\documentclass[12pt,epsfig]{article}
\usepackage{amssymb}
\usepackage{amsfonts}
\usepackage{graphicx}
\usepackage{amsmath}

\input epsf.sty
\newtheorem{theorem}{Theorem}
\newtheorem{acknowledgement}[theorem]{Acknowledgement}

\input{tcilatex}
\makeatletter \@addtoreset{equation}{section}
\renewcommand{\theequation}{\thesection.\arabic{equation}}

\begin{document}

\title{%
\rightline{\mbox {\normalsize
{Lab/UFR-HEP0611/GNPHE/0611/VACBT/0611}}} \textbf{QFT method for indefinite
Kac-Moody Theory: A step towards classification}}
\author{Adil Belhaj$^{1,2,3}$\thanks{%
abelhaj@uottawa.ca}, El Hassan Saidi$^{1,3}$\thanks{%
h-saidi@fsr.ac.ma} \\
{\small $^1$ Groupement National de Physique des Hautes Energies, GNPHE}\\
{\small Siege focal, Lab/UFR-PHE, FSR Rabat, Morocco.}\\
{\small $^2$ Department of Mathematics and Statistics, University of Ottawa}%
\\
{\small 585 King Edward Ave., Ottawa, ON, Canada, K1N 6N5 }\\
{\small $^3$ Virtual African Centre for Basic Science }${\small \&}${\small %
\ Technology}\\
{\small Siege focal, Lab/UFR-PHE, Fac Sciences, Rabat, Morocco.}}
\maketitle

\begin{abstract}
We propose a quantum field theory (QFT) method to approach the
classification of indefinite sector of Kac-Moody algebras. In this approach,
Vinberg relations are interpreted as the discrete version of the QFT$_{2}$
equation of motion of a scalar field and Dynkin diagrams as QFT$_{2}$
Feynman graphs. In particular, we show that Dynkin diagrams of $su\left(
n+1\right) $ series ($n\geq 1$) can be interpreted as free field propagators
and $T_{p,q,r}$.diagrams as the vertex of $\phi ^{3}$ interaction. Other
results are also given.

\textbf{Keywords: }\textit{Vinberg theorem and KM theory, Dynkin diagrams,
QFT Green functions, Feynman graphs.}
\end{abstract}


\thispagestyle{empty}

\newpage \setcounter{page}{1} \newpage

\section{Introduction}

\qquad\ During last decade, the construction of four dimension (4D)
supersymmetric quantum field theories (QFT$_{4}$) has attracted much
attention in 10D superstring theory and D-brane physics $\cite{hw,va}$. It
has been investigated from various points; in particular in type II
superstring models on Calabi-Yau (CY) manifolds with singularities
classified by Dynkin diagrams of Lie algebras $\cite{mayr,r,ca,caa}$. The
physics content of these stringy embedded super- QFTs is obtained from the
deformation of these singularities and the D-branes wraping CY cycles. In
this way, the physical parameters of the QFTs gets a wonderful
interpretation; they are related to the moduli space of CY manifolds with
ADE and conifold geometries $\cite{mayr,ad,adil}$. This result is nicely
obtained in the geometric engineering method by using mirror symmetry in CY
geometries with K3 fibration. The ingredients of the super- QFT$_{4}$
(degrees of freedom, bare masses and gauge coupling constants, RG flows and
cascades, superfields and their group representations,...) are remarkably
encoded in quiver graphs similar to Dynkin diagrams of Kac-Moody (KM)
algebras $\cite{mayr,ad,adil,malika,ma}$.

\qquad These developments have been made possible mainly due to the
correspondence between supersymmetric quiver gauge theories and Dynkin
diagrams of KM algebras. It has been behind the derivation of many
(including exact) results in super- quantum field theories embedded in type
II superstring models. Following $\cite{s}$, the correspondence between
quiver gauge theories and Dynkin diagrams is a powerful tool which can be
made more fruitful in both directions as indicated below:\newline
(\textbf{1}) Use known results on Dynkin diagrams to extract much
information on gauge theory embedded in type II superstring models as
usually done in geometric engineering method:%
\begin{equation*}
Dynkin\text{ }diagrams\qquad \longrightarrow \qquad QFT.
\end{equation*}%
This direction has been extensively explored in literature.\newline
(\textbf{2}) Use standard methods of QFT to complete partial results on
Kac-Moody algebras, in particular their classification and relation to
extraordinary CY singularities beyond ordinary and affine ones:%
\begin{equation*}
QFT\qquad \longrightarrow \qquad Dynkin\text{ }diagrams.
\end{equation*}%
The present study deals with the second direction. Note that at first sight,
this project seems a little bit strange since generally one uses mathematics
to approach physics; but here we are turning the arrow in the other way.
Note also that despite almost four decades since their discovery in 1968,
Kac-Moody extensions of simple Lie algebras $\cite{kac}$ and their
representations have not been fully explored in physics. If forgetting about
unitarity for a while, this disinterest  is also due to the lack of exact
mathematical results with direct relevance for this matter. Only partial
results have been obtained for the so-called KM hyperbolic subset. The
indefinite sector of KM algebras is still an open problem in Lie algebra
theory.

\qquad Motivated by results in type II string theory and its supersymmetric
quiver gauge theory limit, we develop in his paper a QFT method to approach
the classification of Dynkin diagrams of indefinite sector of KM algebras.
Using this method, we show that:\newline
(\textbf{1}) the QFT$_{2}$ equations of motion of a scalar field coincides,
up to discretization, with the statement of Vinberg theorem. The latter is
one of the basic ingredients in KM construction; it gives the classification
of KM algebras into three major subsets. \newline
(\textbf{2}) QFT$_{2}$ Feynamn graphs are interpreted as Dynkin diagrams.
\newline
In addition to above motivations, this field theoretic representation has
moreover direct consequences on the following points: \newline
(\textbf{a}) Shed more light on the striking similarity between Dynkin
diagrams of KM extensions of semi simple Lie algebras and Feynman graphs of
quantum field theory. \newline
(\textbf{b}) Gives a new way to treat the theory of Lie algebras and their
KM classification from physical point of view. \newline
(\textbf{c}) Offers a new method to deal with the KM classification problem
of Dynkin diagrams of indefinite sector of Lie algebras.\newline
(\textbf{d}) Give more insight on the so called indefinite singularities of
CY threefolds encountered in $\cite{malika,malka,laamara}$ and the
corresponding indefinite quiver gauge sector.

\qquad The organization is as follows: In section 2, we review Vinberg
theorem of classification of KM algebras and give the relation with QFT. In
section 3, we propose a two dimensional QFT realization of Vinberg theorem
and KM theory. In section 4, we give the physical representation of Vinberg
condition requiring positivity of Vinberg vectors $\left( u_{i}\right) $.
Last section is devoted to conclusion and discussions.

\section{ \ On Kac-Moody theory: Overview}

\qquad In this section we give an overview on standard KM theory and
preliminary results. Kac-Moody theory is just the extension of semi-simple
Lie algebras of Cartan. The basis of this algebraic construction relies on
the three following:\newline
\textbf{(1)} Vinberg theorem of classification of square matrices $K$; in
particular KM generalized Cartan matrices.\newline
\textbf{(2)} Minimal realization of Vinberg matrices in terms of a triplet.%
\newline
\textbf{(3)} Serre construction of Lie algebras using Chevalley generators.%
\newline
Let us comment briefly these three algebraic steps. Roughly speaking,
Vinberg theorem is a linear algebra theorem which applies to KM theory and
beyond such as Borcherds algebras. This theorem states that the generalized
Cartan matrices $K_{ij}$ (Cartan matrices for short) are of three kinds as
shown here below
\begin{eqnarray}
K_{ij}^{+}u_{j} &>&0,  \notag \\
K_{ij}^{0}u_{j} &=&0,  \label{aa} \\
K_{ij}^{-}u_{j} &<&0.  \notag
\end{eqnarray}%
In these equations, $u_{j}$ are the positive numbers which will be discussed
in section 4. The three upper indices $+$, $-$ and $0$ are conventional
notations introduced in order to distinguish the three KM sectors. The
rigorous statement of Vinberg theorem, as used in KM formulation, is as
follows

\begin{theorem}
\quad A generalized indecomposable Cartan matrix $\mathbf{K}$ obey one and
only one of the following three statements:\newline
(\textbf{1}) \textit{Finite type ( }$\det \mathbf{K}>0$ ): There exist a
real positive definite vector $\mathbf{u}$ ( $u_{i}>0;$ $i=1,2,...$) such
that $\mathbf{K}_{ij}u_{j}=v_{j}>0$.\newline
(\textbf{2}) \textit{Affine type, }corank$\left( \mathbf{K}\right) =1$, $%
\det \mathbf{K}=0$\textit{: }There exist a unique, up to a multiplicative
factor, positive integer definite vector $\mathbf{n}$ ( $n_{i}>0;$ $%
i=1,2,... $) such that $\mathbf{K}_{ij}n_{j}=0$. \newline
(\textbf{3}) \textit{Indefinite type ( }$\det \mathbf{K}\leq 0$ ), corank$%
\left( \mathbf{K}\right) \neq 1$\textit{: } There exist a real positive
definite vector $\mathbf{u}$ ($u_{i}>0;$ $i=1,2,...$) such that $\mathbf{K}%
_{ij}u_{j}=-v_{i}<0$.
\end{theorem}

\quad From the physical point of view, the first sector (ordinary class) of
this KM classification deals with the ordinary semi simple Lie algebras.
These algebras, which are familiar symmetries for model builders of
elementary particle physics, are just the usual finite dimensional algebras
classified many decades ago by Cartan (see figure 1). This model has been
used in $\cite{mayr}$ to describe the geometric engineering of
bi-fundamental matters \bigskip

\begin{figure}[tbh]
\begin{center}
\epsfxsize=7cm \epsffile{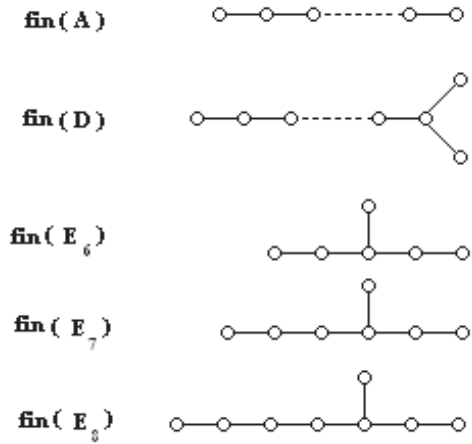}
\end{center}
\caption{{\protect\small \textit{Dynkin diagrams of\ finite dimensional Lie
algebras as classified by Cartan. These are graphs representing the usual }}$%
{\protect\small \mathit{A}}_{n}\sim su\left( n+1\right) ${\protect\small
\textit{\ and }}${\protect\small \mathit{D}}_{n}\sim so\left( 2n\right) $%
{\protect\small \textit{\ classical simple Lie algebras as well as the
ordinary exceptional ones. All of them have symmetric Cartan matrix K }}}
\label{figade}
\end{figure}
The second class (affine class) of KM theory concerns affine Kac-Moody
algebras. The latter plays a basic role in $2d$ conformal field theory (CFT$%
_{2}$) and underlying current algebras. These have been also used in the
geometric engineering of $\mathcal{N}=2$ four dimensional conformal field
theory embedded in Type II superstrings \cite{laamara}. These infinite
dimensional algebras were classified by Kac and Moody; see also figure
2.\bigskip

\begin{figure}[tbh]
\begin{center}
\epsfxsize=7cm \epsffile{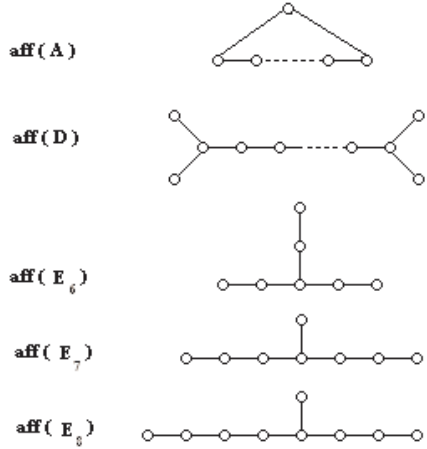}
\end{center}
\caption{\textit{These are Dynkin diagrams of affine Kac-Moody extension of
the corresponding ordinary ones given by figure (\protect\ref{figade}). Like
ordinary graphs, these diagrams are simply laced and }{\protect\small
\textit{symmetric Cartan matrix K \ All these matrices have a vanishing
determinant.}}}
\label{affADE}
\end{figure}
The third class (indefinite class) is the so-called KM indefinite class. In
this sector, we dispose of partial results only; in particular for
hyperbolic subset \cite{malka, sa, vafa, laamara, julia,saidi}, see also
\cite{DHHK, HKN,MMY,H}.

\qquad Before going ahead, let us make two comments regarding the Vinberg
relations (\ref{aa}). First, note that Vinberg relations as shown on \textit{%
theorem 1}, are given by inequalities. However, they can be formulated as
equations by introducing positive quantities $v_{i}$ (vectors) as follows:
\begin{equation}
K_{ij}^{\left( q\right) }u_{j}=qv_{i},\qquad q=+1,0,-1,
\end{equation}%
where the $\left( u_{i}\right) $s and $\left( v_{i}\right) $s are positive
vectors. The second comment we want to make is that, because of the fact
that any irreducible generalized Cartan matrix $K_{ij}^{\left( q\right) }$
can be decomposed as $A_{ij}-\delta A_{ij}^{\left( q\right) }$ with $\delta
A_{ij}^{\left( q\right) }>0$, i.e
\begin{equation}
K_{ij}^{\left( q\right) }=A_{ij}-\delta A_{ij}^{\left( q\right) },
\end{equation}%
the above system of Vinberg equations may be also put in the following
equivalent form
\begin{equation}
A_{ij}u_{j}=w_{i}\left( u\right) ,\qquad A_{ij}=\left( 2\delta _{ij}-\delta
_{i,j+1}-\delta _{i,j-1}\right) ,  \label{12}
\end{equation}%
where appears, on the left side, the ordinary $su\left( n\right) $ Cartan
matrix $A_{ij}$ and where $w_{i}$ are some numbers whose physical meaning
will be given when we consider our QFT realization.

\qquad Concerning the two other points (\textbf{2}) and (\textbf{3}) dealing
with the algebraic construction of KM theory, the key idea of their content
could be summarized as follows. Given a generalized Cartan matrix $K$, one
can associate to it a KM algebra $g\left( K\right) $. This is achieved in
two steps. First by using the minimal realization of Cartan matrix $K$ based
on the usual triplet
\begin{equation}
\left( \hbar ,\Pi ,\Pi ^{v}\right) .
\end{equation}%
This triplet involves the following familiar objects: (\textbf{i}) Cartan
subspace $\hbar $ with a bilinear form $<.,.>$ and a dual space $\hbar
^{\ast }$, (\textbf{ii)} the root basis $\Pi =\left\{ a_{i},\text{ \ \ }%
1\leq i\leq n\right\} \subset \hbar ^{\ast }$ and (\textbf{iii}) the coroot
basis $\Pi ^{v}=\left\{ a_{i}^{v},\text{ \ \ }1\leq i\leq n\right\} \subset
\hbar $. In terms of these quantities, the Cartan matrix reads as
\begin{equation}
K_{ij}=<a_{i}^{v},a_{j}>,
\end{equation}%
which reads generally as $K_{ij}=2a_{i}a_{j}/a_{i}^{2}$. More conveniently,
this can be taken as $K_{ij}=a_{i}a_{j}$ for simply laced KM algebras in
which we will be interested in what follows. Note in passing that this
algebraic formulation is not specific for Kac-Moody extension of semi simple
algebras requiring
\begin{eqnarray}
K_{ii} &=&2,  \notag \\
K_{ij} &<&0,\qquad i\neq j, \\
K_{ij} &=&0\qquad \Rightarrow \qquad K_{ji}=0.  \notag
\end{eqnarray}%
It is also valid for matrices beyond KM generalized Cartan ones. For
instance, this above analysis applies as well for the case of Borcherds
algebras using \textit{real} matrices $\left( B_{ij}\right) $ constrained as
\begin{equation}
2\frac{B_{ij}B_{ji}}{B_{ii}}\in \mathbb{Z},\qquad B_{ii}\neq 0,\qquad
B_{ij}\in \mathbb{R},
\end{equation}%
where $\mathbb{Z}$ is the set of integers. The third step in building KM
algebra $g\left( K\right) $ is based on Chevalley generators $\left\{
e_{i}\right\} $ and $\left\{ f_{i}\right\} $, $i=1,..,n$. The Commutation
relations of KM algebra $g\left( K\right) $ associated with a generalized
Cartan matrix $K$ reads as follows
\begin{equation}
\begin{array}{l}
\left[ e_{i},f_{j}\right] =\delta _{ij}a_{i}^{\nu },\qquad 1\leq i,j\leq n
\\
\left[ h,h^{\prime }\right] =0,\qquad h,h^{\prime }\in \hbar \\
\left[ a_{i}^{\nu },e_{j}\right] =K_{ij}e_{i}, \\
\left[ a_{i}^{\nu },f_{j}\right] =-K_{ij}f_{i}\text{,}%
\end{array}%
\end{equation}%
together with Serre relations. In what follows, we shall develop a quantum
field theoretical method to approach Vinberg theorem and KM theory
describing the extension of semi simple Lie algebras. Our interest into this
quantum field realization is motivated by a set of observations. Here, we
list some of them:\newline
(\textbf{a}) Dynkin diagrams of KM algebras have a remarkable similarity
with the QFT Feynman graphs. For instance, Dynkin diagram of $A_{n}\simeq
su\left( n+1\right) $ semi simple Lie algebra can be interpreted as a scalar
QFT propagator. A naive correspondence reveals that the remaining known
Dynkin diagrams are associated with a special class of QFT Green functions.
It turns out that the Dynkin diagrams of less familiar KM algebras such as
\begin{equation*}
T_{p,q,r}
\end{equation*}%
hyperbolic algebras, with $p,q$ and $r$ positive integers greater than $2,$
have also a QFT counterpart. In particular, the $T_{p,q,r}$s (resp. $%
T_{p_{1},p_{2},p_{3},p_{4}}$) are formally analogous to the three (four)
points tree vertex of scalar quantum field theory with a cubic (quartic)
interaction.\newline
(\textbf{b}) Cartan matrix $A$ of generic $su\left( n+1\right) $ algebras,
with its very particular entries
\begin{equation}
A_{ij}=2\delta _{ij}-\delta _{i,j+1}-\delta _{i,j-1},
\end{equation}%
admits a special factorization, $A=P^{\dagger }P$. It turns out that its
properties are quite similar to those of the $\left( 1+1\right) $
dimensional Laplacian
\begin{equation}
\Delta =\frac{\partial ^{2}}{\partial t^{2}}-\frac{\partial ^{2}}{\partial
x^{2}}=\partial _{+}\partial _{-}
\end{equation}%
of two dimensional QFT ( QFT$_{1+1}$). As we will see later, the $\left(
A_{ij}\right) $ operator is noting but the discrete version of the Laplacian
$\Delta $.\newline
(\textbf{c}) The basis of classification of KM theory rests on Vinberg
theorem relations namely $K_{ij}^{\left( +\right) }u_{j}>0,$ $K_{ij}^{\left(
0\right) }u_{j}=0,\ K_{ij}^{\left( -\right) }u_{j}<0$ where the $K_{ij}$s
are the KM generalized Cartan matrices. These relations, which can be also
put in the compact form
\begin{equation}
K\left( z_{i},z_{j}\right) u\left( z_{j}\right) =v\left( z_{i}\right) ,
\end{equation}%
can be interpreted as quantum field equations of motion obtained from an
action principle. Moreover, in a continuous scalar field $\Phi \left(
t,x\right) $ interpretation, the right term $v\left( z_{i}\right) $ of above
equation would be associated with $\frac{\partial W\left( \Phi \right) }{%
\partial \Phi \left( t,x\right) }$ evaluated at point $z_{i}$. Here $W\left(
\Phi \right) $ is the interacting field potential. In this continuous QFT
limit of Vinberg equations, one also sees that KM affine sector is
associated with the critical points of the field potential $W\left( \Phi
\right) .$ This feature is in agreement with the general picture that we
have about realization of KM affine symmetries and conformal invariance \`{a}
la Sugawara.

\section{QFT representation of Dynkin diagrams}

\qquad To start note that a quantum field realization of Vinberg theorem can
be naturally built by thinking about eq(\ref{12}) as a $\left( 1+1\right) $
dimensional field equation of motion resulting from the variation of the
following discrete field action
\begin{equation}
\mathcal{S}\left[ u\right] =\sum_{i,j\in \mathbb{Z}}\frac{1}{2}%
u_{i}A_{ij}u_{j}+\sum_{i\in \mathbb{Z}}W\left( u_{i}\right) .  \label{sa}
\end{equation}%
In this relation $u_{i}$ is as before, $A_{ij}=\left( 2\delta _{ij}-\delta
_{i,j+1}-\delta _{i,j-1}\right) $ and $W\left( u\right) $ is an interacting
polynomial potential whose variation with respect to $u_{i}$ reads as
follows
\begin{equation}
\frac{\partial W\left( u\right) }{\partial u_{i}}=w_{i}\left( u\right) ,
\end{equation}%
in agreement with eq(\ref{12}). With this discrete field action at hand, one
can go ahead and study quantization of this QFT by computing the generating
functional $\mathcal{Z}\left[ J\right] $ of Green functions of this theory,%
\begin{equation}
\mathcal{Z}\left[ J\right] =\int \left[ Du\right] \exp \left( -\mathcal{S}%
\left[ u\right] -\sum_{i}u_{i}J_{i}\right) .  \label{zj}
\end{equation}%
In this relation $\mathcal{S}\left[ u\right] $ is as in eq(\ref{sa}) and the
$J_{i}$s are the discrete values of an external source dual to the $u_{i}$s.
The two points Green function (propagator) $G_{ij}=<u_{i},u_{j}>$ with $%
\left\vert i-j\right\vert =n,$ is interpreted as the Dynkin diagram of the $%
su\left( n+1\right) $ semi simple Lie algebra; see also figure 3.\bigskip

\begin{figure}[tbh]
\begin{center}
\epsfxsize=7cm \epsffile{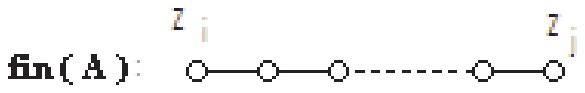}
\end{center}
\caption{A generic {\protect\small \textit{Dynkin diagram of semi simple }A}$%
_{n}$ Lie algebra realized as the two point function $<\Phi \left( z_{i},%
\overline{z}_{i}\right) \Phi \left( z_{j},\overline{z}_{j}\right) >$ with $%
n=1+\left\vert i-j\right\vert $.}
\label{figa}
\end{figure}
More generally, Feynman graphs of the QFT eq(\ref{zj}) should be associated
with Dynkin diagrams. We will not develop here the study of Green functions.
What we want to do now is to establish the general setting of the QFT
realization of KM theory and its relationship with $\left( 1+1\right) $
dimensional continuous quantum scalar field theory.

\begin{theorem}
\qquad The Cartan matrix operator $A_{ij}=\left( 2\delta _{ij}-\delta
_{i,j+1}-\delta _{i,j-1}\right) $ of $su\left( n\right) $ semi simple Lie
algebra is, up a multiplicative constant, exactly equal to the discrete
version of the one dimensional laplacian operator $\Delta =\frac{d^{2}}{%
dx^{2}}$%
\begin{equation}
\Delta \leftrightarrow \frac{1}{a^{2}}A_{ij},
\end{equation}%
where $a$ is period length of the discretized one dimensional lattice.
\newline
Vinberg theorem has a $\left( 1+1\right) $\ QFT realization; and Vinberg
relations ($K_{ij}^{\left( +\right) }u_{j}>0,$ $K_{ij}^{\left( 0\right)
}u_{j}=0,\ K_{ij}^{\left( -\right) }u_{j}<0$) are given by the
discretization of interacting field equations of motion, $A_{ij}u_{j}=\frac{%
\partial W\left( u\right) }{\partial u_{i}}$, with $\partial _{i}W\left(
u\right) >0,$ $\partial _{i}W\left( u\right) =0$ and $\partial _{i}W\left(
u\right) <0$ respectively
\end{theorem}

\qquad Before proving this theorem, let us introduce some tools and useful
convention notations for our QFT realization of KM theory. First, let $\Psi
\left( t,x\right) $ be a $\left( 1+1\right) $ real scalar field of kinetic
energy density
\begin{equation}
\mathcal{E}_{c}=\frac{\partial ^{2}\Psi }{\partial t^{2}}-\frac{\partial
^{2}\Psi }{\partial x^{2}}=\partial _{-}\partial _{+}\Psi .  \label{en}
\end{equation}%
Let also $\mathcal{R}\left( x\right) $ be a static \textit{real positive
definite} scalar field ($\mathcal{R}>0$ and $\frac{\partial \mathcal{R}}{%
\partial t}=0$) varying on the one dimensional real line $\mathbb{R}$.
Because of stationarity, its kinetic energy density, given by a relation
similar to the above one, reduces now to $\mathcal{E}_{c}=-\frac{d^{2}%
\mathcal{R}}{dx^{2}}$. In presence of field interactions $W\left( \mathcal{R}%
\right) $, the action $S=S\left[ \mathcal{R}\right] $ of the scalar field
model is given by
\begin{equation}
S\left[ \mathcal{R}\right] =-\int_{\mathbb{R}}dx\left( \frac{1}{2}\left(
\frac{d\mathcal{R}}{dx}\right) ^{2}+W\left( \mathcal{R}\right) \right) .
\label{ss}
\end{equation}%
The continuous equation of motion of the real positive scalar field $%
\mathcal{R}$ reads as
\begin{equation}
\frac{d^{2}\mathcal{R}}{dx^{2}}=\frac{dW}{d\mathcal{R}},\qquad W\left(
\mathcal{R}\right) =\sum_{m=1}^{n}\lambda _{m}\mathcal{R}^{m},
\end{equation}%
where $\lambda _{m}$\ are coupling constants. To get the discrete version of
this field equation, we use the correspondence $x\rightarrow x_{i}$ and $%
x+dx\rightarrow x_{i}+a$ and denote by
\begin{equation}
\mathcal{R}_{k}=\mathcal{R}\left( x\right) |_{x=x_{k}},\qquad k\in \mathbb{Z}%
,
\end{equation}%
which is nothing but the field value at the node $x_{k}=ka$ of the one
dimensional lattice $\mathbb{Z}$ with $a$ being the lattice period length.%
\newline
We are now in position to prove our theorem. First, consider the discrete
version of energy density $\left( \frac{d\mathcal{R}}{dx}\right) ^{2}$. This
is obtained by help of the usual definition of differentiation namely $\frac{%
d\mathcal{R}\left( x\right) }{dx}=\frac{\mathcal{R}\left( x+dx\right) -%
\mathcal{R}\left( x\right) }{dx}$ and by making the following substitutions
\begin{equation}
\mathcal{R}\left( x\right) \rightarrow \mathcal{R}_{i}\qquad \mathcal{R}%
\left( x+dx\right) \rightarrow \mathcal{R}_{i+1}.
\end{equation}%
Putting these expressions back into the continuous integral $\int_{\mathbb{R}%
}dx\left( \frac{d\mathcal{R}}{dx}\right) ^{2}$, we get the discrete sum $%
\sum_{i\in \mathbb{Z}}\left( \mathcal{R}_{i+1}-\mathcal{R}_{i}\right) ^{2}$
which expands as
\begin{equation}
\sum_{i\in \mathbb{Z}}\left( \mathcal{R}_{i+1}^{2}-\mathcal{R}_{i+1}\mathcal{%
R}_{i}\right) +\sum_{i\in \mathbb{Z}}\left( \mathcal{R}_{i}^{2}-\mathcal{R}%
_{i+1}\mathcal{R}_{i}\right) .
\end{equation}%
Using translation invariance of the one dimensional lattice $\mathbb{Z}$, we
can rewrite the first term of above equation $\sum_{i\in \mathbb{Z}}\left(
\mathcal{R}_{i+1}^{2}-\mathcal{R}_{i+1}\mathcal{R}_{i}\right) $ as
\begin{equation}
\sum_{i\in \mathbb{Z}}\left( \mathcal{R}_{i}^{2}-\mathcal{R}_{i}\mathcal{R}%
_{i-1}\right) .
\end{equation}
This is achieved by shifting the indices as $\left( i+1\right) \rightarrow i$%
. The term $\sum_{i\in \mathbb{Z}}\left( \mathcal{R}_{i+1}-\mathcal{R}%
_{i}\right) ^{2}$ reads then as $\sum_{i\in \mathbb{Z}}\left( 2\mathcal{R}%
_{i}^{2}-\mathcal{R}_{i}\mathcal{R}_{i-1}-\mathcal{R}_{i+1}\mathcal{R}%
_{i}\right) $ and consequently we have the following continuous-discrete
correspondence
\begin{equation}
\frac{1}{2}\int_{\mathbb{R}}dx\left( \frac{d\mathcal{R}}{dx}\right)
^{2}\rightarrow \frac{1}{2a}\sum_{i,j\in \mathbb{Z}}\mathcal{R}_{i}A_{ij}%
\mathcal{R}_{j},
\end{equation}%
where $A_{ij}$ is exactly as given in theorem 2. The presence of the global
factor $\frac{1}{a}$ in front of the discrete sum may be also predicted by
using the following scaling properties of the scalar QFT under change $%
x\rightarrow ax$. In this way, we have
\begin{equation}
\mathcal{R}\left( x\right) \rightarrow \mathcal{R}\left( ax\right) =\mathcal{%
R}\left( x\right) ,\qquad W\left( \mathcal{R}\left( ax\right) \right) =\frac{%
1}{a^{2}}W\left( \mathcal{R}\left( x\right) \right)
\end{equation}%
This completes the proof of our theorem. What remains to do is to find the
physical interpretation of the positivity condition of the $u_{i}$s in
Vinberg theorem. This will be done in the next section.

\section{Vinberg relations as field eq of motion}

\qquad In Vinberg classification theorem of KM algebras (theorem 1), the $%
\left( u_{i}\right) $ variables eq(\ref{aa}) are required to be positive
numbers. From physical point of view, such kind of conditions are familiar
in the study of constrained systems; in particular in gauge theories. In the
problem at hand, Vinberg condition may implemented by considering a static
complex scalar QFT with a $U\left( 1\right) $ gauge symmetry. To do so
consider a QFT system composed by a static one dimensional gauge field $%
\mathcal{A}\left( x\right) $ ( a pure gauge field) and a complex scalar
field $\Phi $
\begin{equation}
\Phi \left( x\right) =\frac{1}{\sqrt{2}}\left[ \Phi _{1}\left( x\right)
+i\Phi _{2}\left( x\right) \right] .
\end{equation}%
For convenience, it is interesting to rewrite the field $\Phi $ by using
Euler representation $\mathcal{R}\left( x\right) \exp i\vartheta \left(
x\right) $ where the field $\mathcal{R}$ is same as before. Using the
following $U\left( 1\right) $ gauge transformations
\begin{eqnarray}
\mathcal{R}\left( x\right) &\rightarrow &\mathcal{R}\left( x\right) ,  \notag
\\
\vartheta \left( x\right) &\rightarrow &\vartheta \left( x\right) -\lambda
\left( x\right) , \\
\mathcal{A}\left( x\right) &\rightarrow &\mathcal{A}\left( x\right) -i\frac{%
d\lambda \left( x\right) }{dx}  \notag
\end{eqnarray}%
where $\lambda \left( x\right) $ is the gauge parameter, and the gauge
covariant derivative $\mathcal{D}=\frac{d}{dx}+i\mathcal{A}\left( x\right) $%
, one can write down the static one dimensional action $S\left[ \Phi \right]
$ describing the complex scalar field dynamics. It reads as,%
\begin{equation}
S\left[ \Phi \right] =-\int_{\mathbb{R}}dx\left[ \left( \mathcal{D}\Phi
\right) ^{\ast }\left( \mathcal{D}\Phi \right) +W\left( \left\vert \Phi
\right\vert \right) \right] ,
\end{equation}%
where $W\left( \left\vert \Phi \right\vert \right) =W\left( \mathcal{R}%
\right) $ is gauge invariant interacting potential, the same as in eq(\ref%
{ss}). Using gauge symmetry of this action, one can make the gauge choice
\begin{equation}
\vartheta \left( x\right) =\lambda \left( x\right) ,\qquad \mathcal{A}\left(
x\right) =i\frac{d\lambda \left( x\right) }{dx},\qquad \mathcal{D}\Phi =%
\frac{d\mathcal{R}}{dx},  \label{ga}
\end{equation}%
to kill the local phase $\vartheta \left( x\right) $ of the complex field $%
\Phi \left( x\right) $ which reduces then to $\mathcal{R}\left( x\right) $.
Vinberg condition corresponds then to fixing the gauge field.

\section{Conclusion and discussion}

\qquad In this paper, we have developed the basis of a quantum field
realization of KM theory of Lie algebras. As we know this structure, encoded
by the Dynkin diagrams, play a central role in quantum physics and has been
behind the developments of gauge theory and 2D critical phenomena.

\qquad In the case of simply laced Dynkin diagrams, we have shown that
Vinberg theorem, classifying KM algebras, is in fact just the discrete
version of the static field equation of motion
\begin{equation}
\frac{d^{2}\mathcal{R}}{dx^{2}}=\frac{dW\left( \mathcal{R}\right) }{d%
\mathcal{R}},\qquad \Phi =\mathcal{R}\exp i\vartheta ,
\end{equation}%
following from the minimization of a complex scalar $U\left( 1\right) $
gauge invariant theory. Gauge symmetry is used to fix the phase $\vartheta $
of the field $\Phi $ and the original field action $\mathcal{S}\left[
\mathcal{R},\vartheta ,\mathcal{A}\right] $ is left with only a dependence
in the positive field $\mathcal{R}$. In this approach, Vinberg condition
requiring positivity of the $u_{i}$s is interpreted as corresponding to the
gauge fixing of $U\left( 1\right) $ invariance eq(\ref{ga}). According to
the sign of $\frac{dW}{d\mathcal{R}}$, one distinguishes then three sectors,
\begin{eqnarray}
\frac{dW}{d\mathcal{R}} &>&0,\qquad  \notag \\
\frac{dW}{d\mathcal{R}} &=&0,\qquad \\
\frac{dW}{d\mathcal{R}} &<&0.  \notag
\end{eqnarray}%
In this representation, one sees that affine KM sector is associated with
the critical point of the interacting field potential $W\left( \mathcal{R}%
\right) $ ($\frac{dW}{d\mathcal{R}}=0$). Semi simple Lie algebras are
associated with%
\begin{equation}
\frac{dW}{d\mathcal{R}}>0,
\end{equation}%
and interpreted as stable fluctuations around the critical point while
indefinite symmetries related with unstable deformations,%
\begin{equation}
\frac{dW}{d\mathcal{R}}<0.
\end{equation}%
In a subsequent study $\cite{as}$, we give other applications and more
explicit details on this construction; in particular on the generating
functional $\mathcal{Z}$ of Dynkin diagrams of KM algebras.

\begin{acknowledgement}
\qquad This research work is supported by the program protars III, CNRST
D12/25. We thank R. Ahl Laamara, M. Ait Benhaddou and L.B Drissi for earlier
collaboration on this matter.
\end{acknowledgement}

\end{document}